\documentclass[sigconf]{acmart}
\usepackage[most]{tcolorbox}
\usepackage{xcolor}
\usepackage{enumitem}
\usepackage{longtable}
\usepackage{booktabs}

\AtBeginDocument{%
  \providecommand\BibTeX{{%
    \normalfont B\kern-0.5em{\scshape i\kern-0.25em b}\kern-0.8em\TeX}}}

\setcopyright{acmlicensed}
\copyrightyear{2026}
\acmYear{2026}

\acmDOI{XXXXXXX.XXXXXXX}

\acmConference[Conference '26]{Conference Full Name}{June 03--05, 2026}{Woodstock, NY}
\acmBooktitle{Conference Full Name (Conference '26), June 03--05, 2026, Woodstock, NY}

\acmISBN{978-1-4503-XXXX-X/26/06}

\graphicspath{{./images/}}

\begin{document}

\title{Auditing Student–AI Collaboration: A Case Study of Online Graduate CS Students}

\author{Nifu Dan}
\email{ndan3@gatech.edu}
\orcid{0000-0000-0000-0000}
\affiliation{%
  \institution{Georgia Institute of Technology}
  \streetaddress{North Avenue}
  \city{Atlanta}
  \state{Georgia}
  \country{USA}
  \postcode{30332}
}

\renewcommand{\shortauthors}{Dan, N.}

\begin{abstract}
As generative AI becomes embedded in higher education, it increasingly shapes how students complete academic tasks. While these systems offer efficiency and support, concerns persist regarding over-automation, diminished student agency, and the potential for unreliable or hallucinated outputs. This study conducts a mixed-methods audit of student–AI collaboration preferences by examining the alignment between current AI capabilities and students’ desired levels of automation in academic work. Using two sequential and complementary surveys, we capture students’ perceived benefits, risks, and preferred boundaries when using AI. The first survey employs an existing auditing framework to assess preferences for and actual usage of AI across 12 tasks in educational contexts, alongside primary concerns and reasons for use. The second survey, informed by the first, explores how AI systems could be designed to address these concerns through open-ended questions. This study aims to identify gaps between existing AI affordances and students’ normative expectations of collaboration, informing the development of more effective and trustworthy AI systems for education.
\end{abstract}



\maketitle

\section{Introduction}

Generative artificial intelligence (GenAI) tools such as ChatGPT, Claude, and Gemini are rapidly becoming embedded in higher education, reshaping how students complete academic tasks~\cite{kasneci2023chatgpt, chan2024systematic, sukumar2025generative}. These systems promise substantial gains in efficiency, personalization, and support, yet they also introduce significant risks related to over-automation, reduced student agency, and vulnerability to unreliable or hallucinated outputs~\cite{shneiderman2020human, ji2023survey, athaluri2023exploring}. As students increasingly delegate cognitive and procedural work to AI, it becomes essential to understand not only how they use these tools, but also how they perceive the trade-offs between automation and autonomy, and what design features would make AI systems more trustworthy in educational contexts~\cite{khoebel2022explainable, amershi2019guidelines}.

Recent research highlights the dual-edged nature of GenAI in education. Systematic reviews indicate that these tools can enhance writing quality, support personalized learning, and increase productivity, while simultaneously raising concerns about academic integrity, bias, and equitable access~\cite{sukumar2025generative, li2025generative, camacho2025ethical}. Empirical studies further suggest that the impact of AI depends heavily on how it is integrated pedagogically~\cite{lee2024harnessing, maizatul2025implementing}. However, a critical gap remains in understanding how students themselves navigate collaboration with AI across diverse tasks in educational setting.

This study addresses that gap by conducting a mixed-methods audit of student–AI collaboration among graduate students in the Georgia Tech Online Master of Science in Computer Science (OMSCS) program. Grounded in a Human-Centered AI (HCAI) perspective~\cite{shneiderman2020human, shneiderman2022human} and informed by the Human Agency Scale (HAS)~\cite{shao2025future}, we investigate two core dimensions of this collaboration. First, we measure students' preferred levels of automation across twelve common tasks in educational contexts and compare these preferences to their actual AI usage~\cite{shao2025future, shao2024collaborative}. This allows us to map tasks into four alignment zones: Green Light (high desire, high use), R\&D Opportunity (high desire, low use), Low Priority (low desire, low use), and Red Light (low desire, high use). Second, through qualitative analysis of open-ended responses, we identify the system-level features that students believe would enhance the trustworthiness of AI in education~\cite{amershi2019guidelines, khoebel2022explainable, khosravi2024trustworthy}.\\
\indent In this paper, we investigate the following questions:
\begin{itemize}[noitemsep, topsep=0pt]
    \item \textbf{RQ1:} What levels of automation do students desire across different academic tasks, and how do these preferences align with their actual AI usage?
    
    \item \textbf{RQ2:} How do students' reasons for using AI and their associated concerns vary across task types, and what does this reveal about the perceived risks and benefits of student--AI collaboration?
    
    \item \textbf{RQ3:} What system features and design principles do students identify as most important for addressing their concerns?
\end{itemize}

\section{Related Work}

\subsection{Generative AI in Higher Education}

A growing body of literature documents the transformative but contested role of generative AI in academic settings. Systematic reviews and meta-analyses highlight that tools such as ChatGPT can improve writing fluency, reduce cognitive load for routine tasks, and support personalized learning experiences when used appropriately \cite{kasneci2023chatgpt}. These benefits are most apparent for surface-level tasks like grammar correction, summarization, and initial drafting, where AI functions as a productivity aid rather than a substitute for learning.

At the same time, substantial concerns have been raised about over-reliance on AI, hallucinated or biased outputs, and the potential erosion of critical thinking and disciplinary epistemic skills \cite{bastani2024generative,selwyn2019should}. Empirical work suggests that when students delegate substantive intellectual work—such as complex problem solving or argumentation—to AI systems without appropriate scaffolding, learning outcomes can suffer \cite{bastani2024generative}.

Instructional context and pedagogical design play a decisive role in mediating these effects. For example, research comparing AI-generated feedback with human or peer feedback finds that AI is effective for addressing surface-level concerns aligned with rubrics, but that peer or instructor feedback remains essential for fostering deeper conceptual understanding and disciplinary reasoning \cite{cho2007scaffolding,liu2023aiwriting}. Such work converges on the conclusion that generative AI is most beneficial when integrated as a complement to human pedagogical interaction, rather than as a replacement.

\subsection{Human-Centered AI in Educational Contexts}

Human-Centered AI (HCAI) has emerged as a foundational paradigm for responding to the risks posed by opaque, overly autonomous AI systems. Rooted in human–computer interaction and AI ethics, HCAI advocates for systems that are reliable, safe, and trustworthy while preserving meaningful human control and agency \cite{shneiderman2020human}. Rather than optimizing for automation alone, HCAI emphasizes augmentation that supports human goals, values, and accountability.

In educational contexts, HCAI principles foreground the importance of preserving both student and instructor agency, enhancing transparency of system behavior, and embedding safeguards against misuse, bias, and automation bias \cite{amershi2019guidelines,holmes2022ethics}. Design recommendations from this literature include making system limitations visible, offering explanations and confidence cues, and enabling users to intervene, override, or verify AI outputs. These features are particularly critical in learning environments where inappropriate trust or over-dependence on AI can undermine motivation, metacognition, and skill development.

\subsection{Human Agency and Automation Preferences}

As AI systems become active collaborators in human workflows, researchers have sought to quantify how users perceive and negotiate their agency under automation. A recent contribution to this literature is the \textbf{Human Agency Scale (HAS)} introduced by Shao et al., which provides a five-level scale (H1–H5) to quantify the degree of human involvement preferred in task performance when working with AI agents \cite{shao2025future}. HAS moves beyond binary automate-or-not frameworks by offering a shared language for examining desired human control versus automation or augmentation across diverse tasks and occupations.

Our study adapts the HAS framework to education domain, applying it at a task-specific level to compare desired human agency and actual AI use. This approach enables a fine-grained audit of student–AI collaboration and provides empirical grounding for human-centered design recommendations.

\section{Methodology}

\begin{table}[t]
    \centering
    \caption{12 tasks in educational contexts used in the study.}
    \label{tab:task-list-12}
    \renewcommand{\arraystretch}{1.15}
    \begin{tabular}{@{}clp{12cm}@{}}
        \toprule
        \textbf{ID} & \textbf{Task Description} \\
        \midrule
        T1  & Summarize a research article. \\
        T2  & Formatting Citation and Bibliography. \\
        T3  & Brainstorm essay ideas or outline structure\\
        T4  & Revise Writing for grammar and style. \\
        T5  & Debug code snippet or explain errors. \\
        T6  & Create personalized study plan. \\
        T7  & Generate flashcards or practice quizzes \\
        T8  & Recommend Learning Resources such as video or textbook \\
        T9  & Draft professional emails to TA or Professor \\
        T10 & Help explain or solve step by step quantitative problem \\
        T11 & Summarize class note or discussion \\
        T12 & Give feedback on draft writing \\
        \bottomrule
    \end{tabular}
\end{table}

\subsection{12 Tasks in Educational Context}

To systematically evaluate student preferences for AI collaboration, we selected twelve tasks in educational setting. Each task was chosen based on its relevance to both conceptual and practical skill development. For instance, tasks such as debugging code (T5) and solving quantitative problems step-by-step (T10) represent core technical competencies, while activities like summarizing research (T1) and providing feedback on writing (T12) emphasize critical engagement and communication skills~\cite{shao2024assisting}. By including tasks that range from routine (e.g., formatting citations) to complex and open-ended (e.g., brainstorming essay structures), we aimed to capture a spectrum of automation preferences and perceived risks.

Table~\ref{tab:task-list-12} presents the complete list of tasks.

\subsection{Desired vs. Actual Automation Likert Questions}

\begin{figure*}[t]
\centering

\begin{minipage}[t]{0.48\linewidth}
\centering
\begin{tcolorbox}[
    colback=white,
    colframe=black,
    boxrule=1pt,
    arc=4pt,
    left=8pt,
    right=8pt,
    top=6pt,
    bottom=6pt,
    width=\linewidth
]
\begin{tcolorbox}[
    colback=black!80,
    colframe=black!80,
    boxrule=0pt,
    arc=3pt,
    left=6pt,
    right=6pt,
    top=4pt,
    bottom=4pt
]
\textcolor{white}{\textbf{Likert Question for Collecting Automation Desire}}
\end{tcolorbox}

\vspace{6pt}

\small
\textbf{Question.}  
For each academic task, please rate the level of automation that an AI system
\emph{should ideally provide to support learning}, from your (a student's) perspective.

\vspace{6pt}

\textbf{Scale.}
\begin{itemize}[leftmargin=1.2em]
    \item \textbf{1:} AI should not get involved in this task at all
    \item \textbf{2:} AI may give minor suggestions
    \item \textbf{3:} AI and human should share responsibility
    \item \textbf{4:} AI should handle most of the task with human oversight
    \item \textbf{5:} AI should perform this task autonomously
\end{itemize}

\end{tcolorbox}

\captionof{figure}{Desired level of AI automation.}
\label{fig:likert-automation}
\end{minipage}
\hfill
\begin{minipage}[t]{0.48\linewidth}
\centering
\begin{tcolorbox}[
    colback=white,
    colframe=black,
    boxrule=1pt,
    arc=4pt,
    left=8pt,
    right=8pt,
    top=6pt,
    bottom=6pt,
    width=\linewidth
]
\begin{tcolorbox}[
    colback=black!80,
    colframe=black!80,
    boxrule=0pt,
    arc=3pt,
    left=6pt,
    right=6pt,
    top=4pt,
    bottom=4pt
]
\textcolor{white}{\textbf{Likert Question for Collecting Observed AI Usage}}
\end{tcolorbox}

\vspace{6pt}

\small
\textbf{Question.}  
For each academic task listed below, please indicate how capable current AI systems are
\emph{based on your own experience}, reflected by how you actually use AI for that task.

\vspace{6pt}

\textbf{Scale.}
\begin{itemize}[leftmargin=1.2em]
    \item \textbf{1:} I do not use AI for this task at all
    \item \textbf{2:} I use AI to provide minor or partial help
    \item \textbf{3:} I use AI for this task moderately
    \item \textbf{4:} I use AI to handle most of the task with oversight
    \item \textbf{5:} I rely entirely on AI to automate this task
\end{itemize}

\end{tcolorbox}

\captionof{figure}{Actual AI usage Level.}
\label{fig:likert-observed}
\end{minipage}

\end{figure*}

Following prior work by Shao et al.~\cite{shao2025future} on characterizing levels of AI automation, we operationalize students' automation preferences using a five-point Likert-scale that reflects a continuum of responsibility allocation between humans and AI (Figure~\ref{fig:likert-automation}). Rather than treating automation as a binary choice, this scale explicitly distinguishes between AI serving a purely assistive role, sharing responsibility with the human learner, or assuming primary or full control over the task~\cite{shao2024collaborative}. This design allows us to capture nuanced automation boundaries across academic tasks and to examine where students draw limits on AI involvement in educational contexts~\cite{venkatesh2003user}.

In contrast to prior work by Shao et al.~\cite{shao2025future}, which characterizes AI capability using expert annotations, our study does not rely on external expert judgments of system performance. Instead, we capture students' perceived AI capability through their \emph{self-reported actual usage and reliance} on AI systems across academic tasks (Figure~\ref{fig:likert-observed})~\cite{dwivedi2023chatgpt, sullivan2023chatgpt}.

Specifically, we ask participants to indicate how they currently use AI for each task, ranging from not using AI at all to relying entirely on AI to automate the task. This formulation operationalizes perceived capability in terms of habitual use and practical reliance, reflecting students' lived experiences with contemporary AI tools rather than an objective assessment of model correctness or task success~\cite{jian2000foundations}.

\subsection{Four-Zone Classification of AI Automation}

To analyze students' perceptions of appropriate AI involvement in academic tasks, we adopted a four-zone classification of automation adapted from Shao et al.'s framework~\cite{shao2025future} for integrating human worker and AI expert perspectives on automation. This framework partitions the automation landscape into four zones based on the alignment or misalignment between human preferences and perceived AI capability.

The four zones are defined as follows:

\textbf{Automation ``Green Light'' Zone.}
Tasks in the Green Light zone are those for which both humans and AI systems are perceived as well-suited for automation. In this zone, there is strong alignment between participants' willingness to delegate a task to AI and their confidence in AI's ability to perform the task effectively. These tasks represent contexts in which AI adoption is broadly acceptable and likely to provide immediate benefits with minimal resistance~\cite{shao2025future}.

\textbf{Automation ``Red Light'' Zone.}
The Red Light zone includes tasks that participants believe should not be automated, even if AI systems are technically capable of performing them. This zone reflects strong human resistance to AI involvement, often due to concerns related to trust, accountability, ethical implications, or academic integrity~\cite{cotton2023chatting, perkins2023academic}. Tasks in this zone signal boundaries where AI use is perceived as inappropriate despite potential performance gains.

\textbf{R\&D Opportunity Zone.}
Tasks in the R\&D Opportunity zone are those for which participants express openness to AI assistance but lack confidence in current AI systems' capabilities. This zone highlights gaps between user expectations and existing technological performance and points to opportunities for further AI development, evaluation, and design refinement~\cite{shao2025future}. Improving transparency, reliability, or controllability of AI systems may enable tasks in this zone to transition toward the Green Light zone~\cite{kasneci2023chatgpt}.

\textbf{Low Priority Zone.}
The Low Priority zone consists of tasks that participants neither desire to automate nor believe AI systems are well-suited to perform. These tasks are typically viewed as either unimportant for automation or inherently human-centered, resulting in low perceived value of AI involvement. From a design perspective, this zone suggests limited benefit in prioritizing AI development for such tasks~\cite{shao2025future}.

\subsection{Primary Concerns and Reasons for AI Usage across Academic Tasks}

\begin{table*}[t]
\centering
\small
\begin{tabular}{p{0.30\textwidth} p{0.60\textwidth}}
\toprule
\textbf{Category} & \textbf{Survey Item Description} \\
\midrule
\textbf{Primary Concerns} & \\
Inaccurate or misleading information & Concern that AI outputs may contain factual errors or hallucinations that affect task correctness \\
Risk of academic misconduct & Concern about plagiarism, policy violations, or inappropriate AI use in graded work \\
Reduced critical thinking & Concern that relying on AI may weaken students' independent reasoning or learning \\
\midrule
\textbf{Reasons for AI Usage} & \\
Saving time & Using AI to complete tasks more efficiently and reduce time spent on repetitive work \\
Reducing cognitive load & Using AI to manage mentally demanding or complex tasks \\
Improving output quality & Using AI to enhance clarity, accuracy, or polish of academic outputs \\

\bottomrule
\end{tabular}
\caption{Selection of primary concerns and reasons for AI usage included in the survey.}
\label{tab:concerns-reasons}
\end{table*}

To understand what shapes students' adoption of AI across different academic tasks, we asked participants to report both their primary concerns when using AI and their reasons for relying on AI assistance~\cite{dwivedi2023chatgpt, sullivan2023chatgpt}. Concerns included risks such as inaccurate or misleading outputs, hallucinations, academic misconduct, and reduced critical thinking~\cite{cotton2023chatting, halaweh2023chatgpt, imran2023analyzing}, while reasons for use captured perceived benefits such as saving time, reducing cognitive load while managing conceptually complex task and improving accuracy or polishing academic outputs~\cite{kasneci2023chatgpt}, as shown in Table 2.

\subsection{Design Expectations for Addressing AI Concerns}

In addition to structured Likert-scale items, we included open-ended question to elicit students' perspectives on how AI systems should be designed to address their greatest concern in educational contexts~\cite{braun2006using}. The use of open-ended questions allows participants to express design expectations in their own terms, capturing aspects of AI system behavior and interface design that may not be fully anticipated by closed-form survey items~\cite{braun2006using}. We treat these responses as qualitative inputs for identifying recurring design considerations and informing subsequent analysis of user-centered approaches to AI alignment in education~\cite{swiecki2022assessment}.

\section{User Study Method}
\subsection{Participants}
Participants were graduate students enrolled in the Online Master of Science in Computer Science (OMSCS) program at the Georgia Institute of Technology. Participation in the survey was voluntary. To ensure the relevance of responses to AI-assisted coursework, we filtered out participants who reported rarely using or not using AI tools in their academic work.
In total, $N = 57$ students completed the survey. After filtering, $N = 44$ participants who reported at least occasional use of AI systems
for coursework were retained for analysis.
The final sample therefore consists of students with meaningful experience interacting with AI
in educational contexts, allowing us to focus our analysis on substantive patterns of AI use.

\subsection{Survey Based Interviews}

In Phase 1, participants completed a questionnaire assessing their use of AI tools and their perceptions of appropriate levels of AI automation across a range of tasks in educational contexts.
Participants also reported their primary reasons and concerns regarding AI use for each task. Using the responses from Phase 1, we identified the most frequently reported concern across tasks through quantitative aggregation of participants’ ratings. In Phase 2, we conducted a follow-up survey focused specifically on this most prevalent concern. Participants were asked to describe the AI system features, design principles, or safeguards they believed would effectively address this concern in educational contexts. This second phase enabled us to move from identifying perceived risks to systematically collecting participant-informed design expectations.

\subsection{Ethics}

The study followed established ethical guidelines for research involving human participants.
Prior to participation, all participants completed an informed consent form.
Participation was voluntary, and no personally identifiable information was collected.
All responses were anonymized prior to analysis and used solely for research purposes.

The study posed minimal risk to participants.
All questions focused on participants’ experiences and perceptions of AI use in academic contexts,
and no deception was involved.
Data were collected and stored in a manner that protected participant privacy
and ensured confidentiality.

\subsection{Data Analysis}

To compare students’ desired levels of AI involvement with their reported use of AI tools,
we operationalized two aggregate measures: \emph{aggregate desire level} and
\emph{aggregate usage level}.
Both measures were derived from participants’ Likert-scale responses
and computed at the task level.

\paragraph{Aggregate Desire Level.}
For each academic task $t$, participants indicated the level of AI automation they believed
was appropriate using an ordered Likert-scale corresponding to increasing levels of automation
(from minimal AI involvement to fully autonomous AI execution).
Responses were numerically encoded such that higher values indicate a greater desired level
of AI automation.
The aggregate desire level for task $t$ was calculated as the mean of these encoded responses
across all participants in the analytic sample ($N = 44$):

\[
D_t = \frac{1}{N} \sum_{i=1}^{N} d_{i,t},
\]

where $d_{i,t}$ denotes participant $i$’s desired automation rating for task $t$.

\paragraph{Aggregate Usage Level.}
Actual AI usage was operationalized using participants’ self-reported frequency of AI tool use.
Responses were encoded on an ordered numerical scale, with higher values indicating more
frequent AI use.
The aggregate usage level was computed as the mean usage score across all participants and tasks:
\[
U_t = \frac{1}{N} \sum_{i=1}^{N} u_{i,t}.
\]

\section{User Study Results}

\begin{figure*}[t]
    \centering
    \includegraphics[width=\textwidth,,height=0.35\textheight]{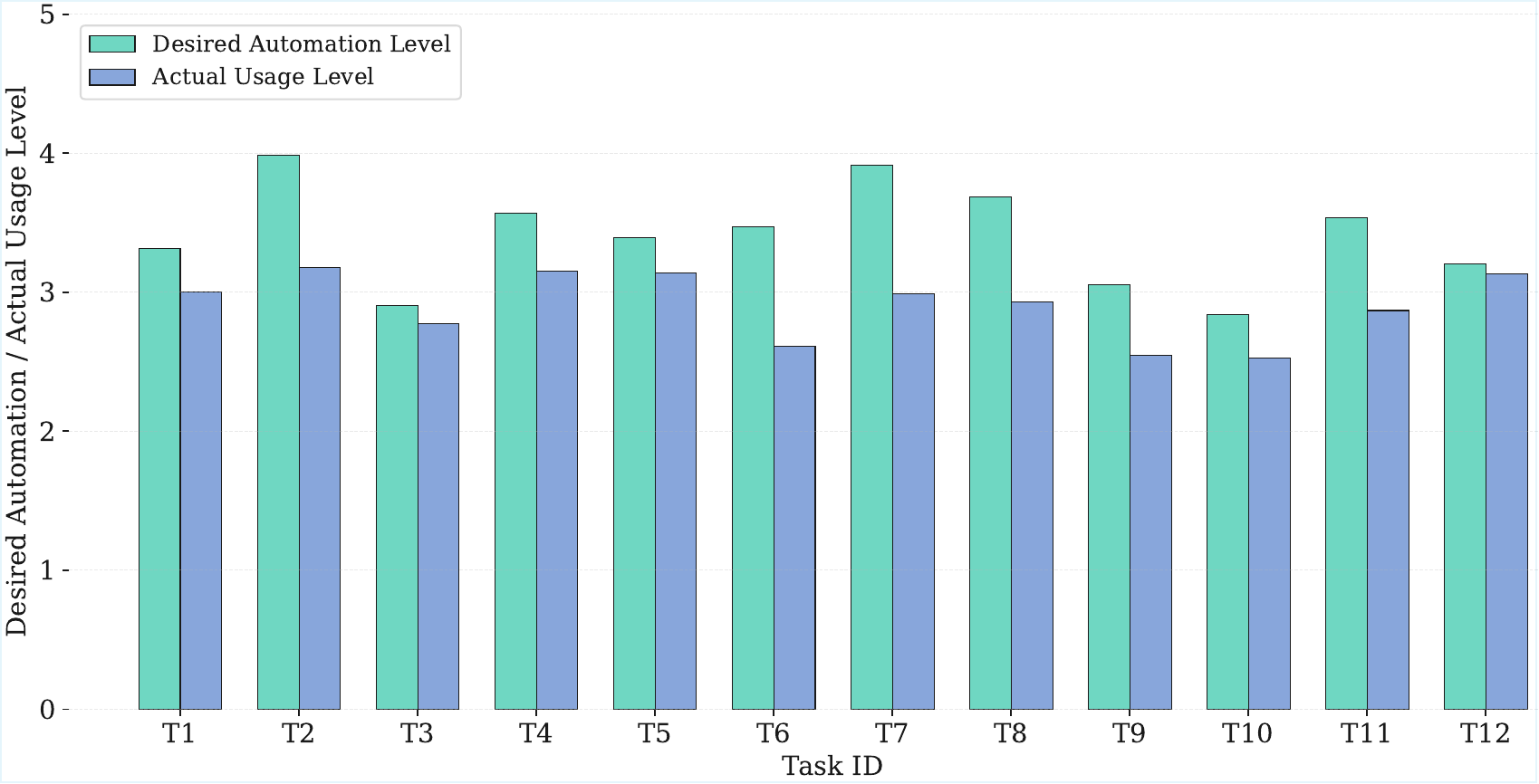}
    \caption{Comparison of students’ \emph{desired AI automation levels} and \emph{self-reported actual AI usage} across twelve tasks (T1–T12).}
    \label{fig:placeholder}
\end{figure*}

\begin{figure}
    \centering
    \includegraphics[width=1\linewidth,,height=0.28\textheight]{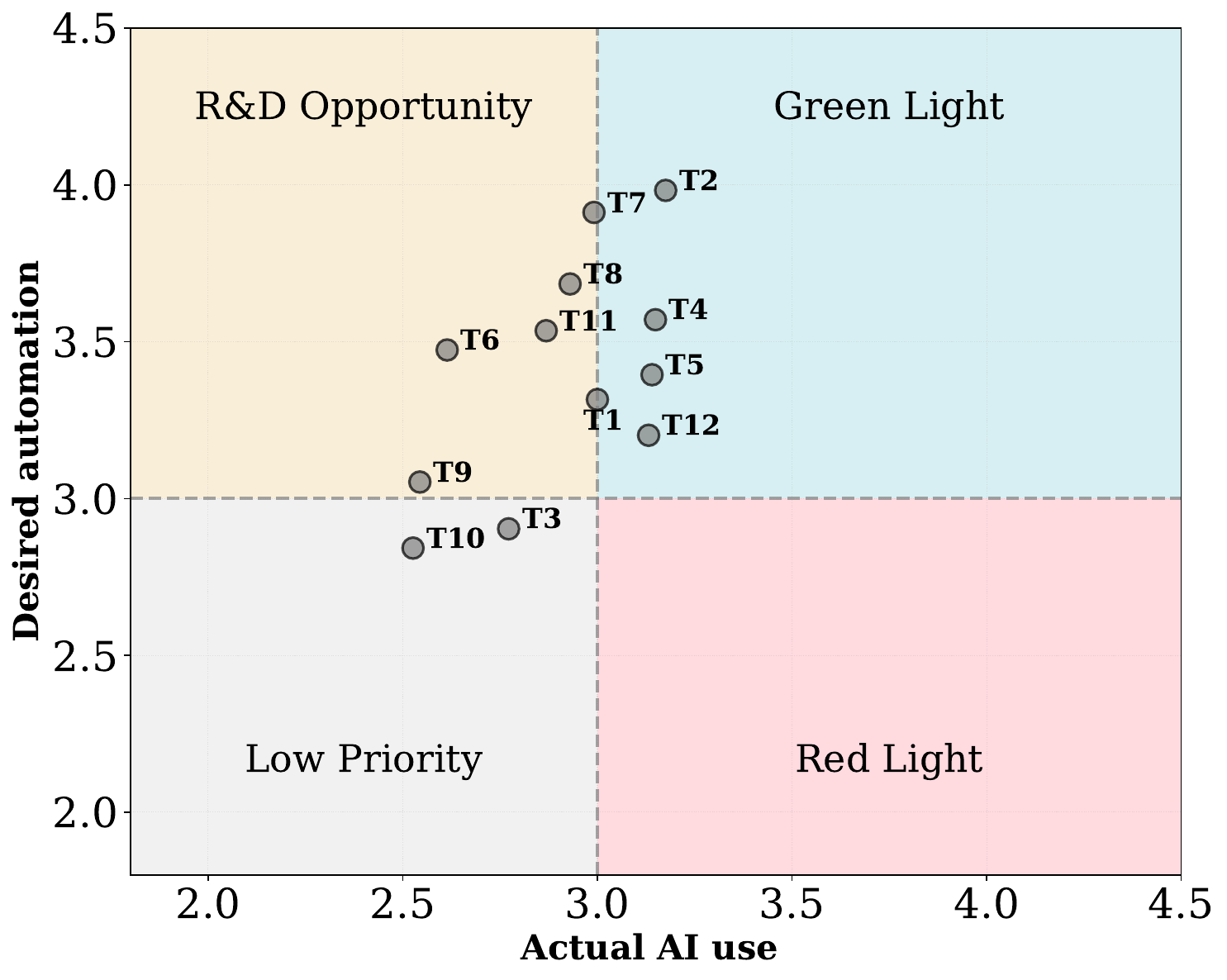}
    \caption{Four-zone automation alignment map plotting desired AI automation against actual AI usage for twelve tasks.}
    \label{fig:placeholder2}
\end{figure}

\subsection{RQ1: Desired Automation Level vs. Actual AI Usage Level}

Figure~\ref{fig:placeholder} illustrates participants’ aggregate desired automation levels
and reported AI usage levels across twelve academic tasks (T1–T12).
Across most tasks, desired automation levels exceed reported usage,
indicating a consistent gap between participants’ preferences for AI involvement
and their current patterns of AI use.
However, the magnitude of this gap varies substantially by task type,
suggesting that task characteristics play an important role in shaping how students
engage with AI tools in practice.

\subsubsection{Information Compression and Retrieval Tasks}
Summarization and information organization (T1, T11) show relatively high desired automation and moderately high usage. 
These largely extractive, low-reasoning tasks are perceived as supportive rather than generative. 
Because errors are easier to detect and consequences are limited, participants are comfortable using AI, resulting in a relatively small desire–usage gap.

\subsubsection{Writing Support and Surface-Level Editing Tasks}
Grammar revision (T4), citation formatting (T2), and draft feedback (T12) exhibit some of the highest desired automation levels. 
As procedural and rule-based activities, they are well suited to AI assistance. 
However, actual usage remains lower than desired, suggesting caution related to correctness, formatting standards, and academic norms.

\subsubsection{Ideation and Study Support Tasks}
 Study planning (T6), flashcard generation (T7), and resource recommendation (T8) display some of the largest desire–usage gaps. 
Although participants express strong interest in AI support, usage is comparatively low. 
These tasks require contextualization and personalization, areas where AI is perceived as inconsistently reliable, producing openness tempered by reservations.

\subsubsection{Communication and Socially Sensitive Tasks}
Professional email drafting (T9) shows moderate desired automation but relatively low usage. 
Here, the gap appears driven less by perceived capability and more by interpersonal concerns. 
Participants may hesitate to delegate tasks involving tone, professionalism, and accountability.

\subsubsection{Analytical and High-Stakes Reasoning Tasks}
Analytical tasks vary. 
For code debugging (T5), usage closely aligns with desired automation, reflecting confidence in outputs that can be tested. 
In contrast, quantitative problem solving (T10) shows a larger gap. 
Despite moderate interest in AI support, participants report lower usage, likely due to the precision required and the difficulty of detecting reasoning errors.

\subsubsection{ Four Automation Zones}

Figure~\ref{fig:placeholder2} shows that the twelve academic tasks
are distributed unevenly across the four automation zones,
revealing systematic patterns in how students’ preferences for AI automation
align with their reported usage.

Several tasks cluster in the region characterized by both relatively high desired automation
and high actual AI usage.
These include formatting citations and bibliographies,
revising writing for grammar and style,
summarizing class notes or discussions,
summarizing research articles,
and debugging code.
The concentration of tasks in this region suggests that students are already
actively using AI for tasks that are procedural, supportive,
or whose outputs can be readily verified.

A second group of tasks falls into a region where desired automation is high
but actual usage remains comparatively lower.
This group includes creating personalized study plans,
generating flashcards or practice quizzes,
recommending learning resources,
drafting professional emails,
and solving step-by-step quantitative problems.
The placement of these tasks indicates that while students express interest
in greater AI involvement,
their current usage lags behind their preferences,
highlighting areas where existing AI tools may not yet fully meet students’
expectations or where contextual concerns limit adoption.

\subsection{RQ2: Primary Reasons and Concerns of using AI across tasks}

\begin{figure*}[t]
\centering
\begin{minipage}[t]{0.48\textwidth}
    \centering
    \includegraphics[width=\linewidth,height=0.29\textheight]{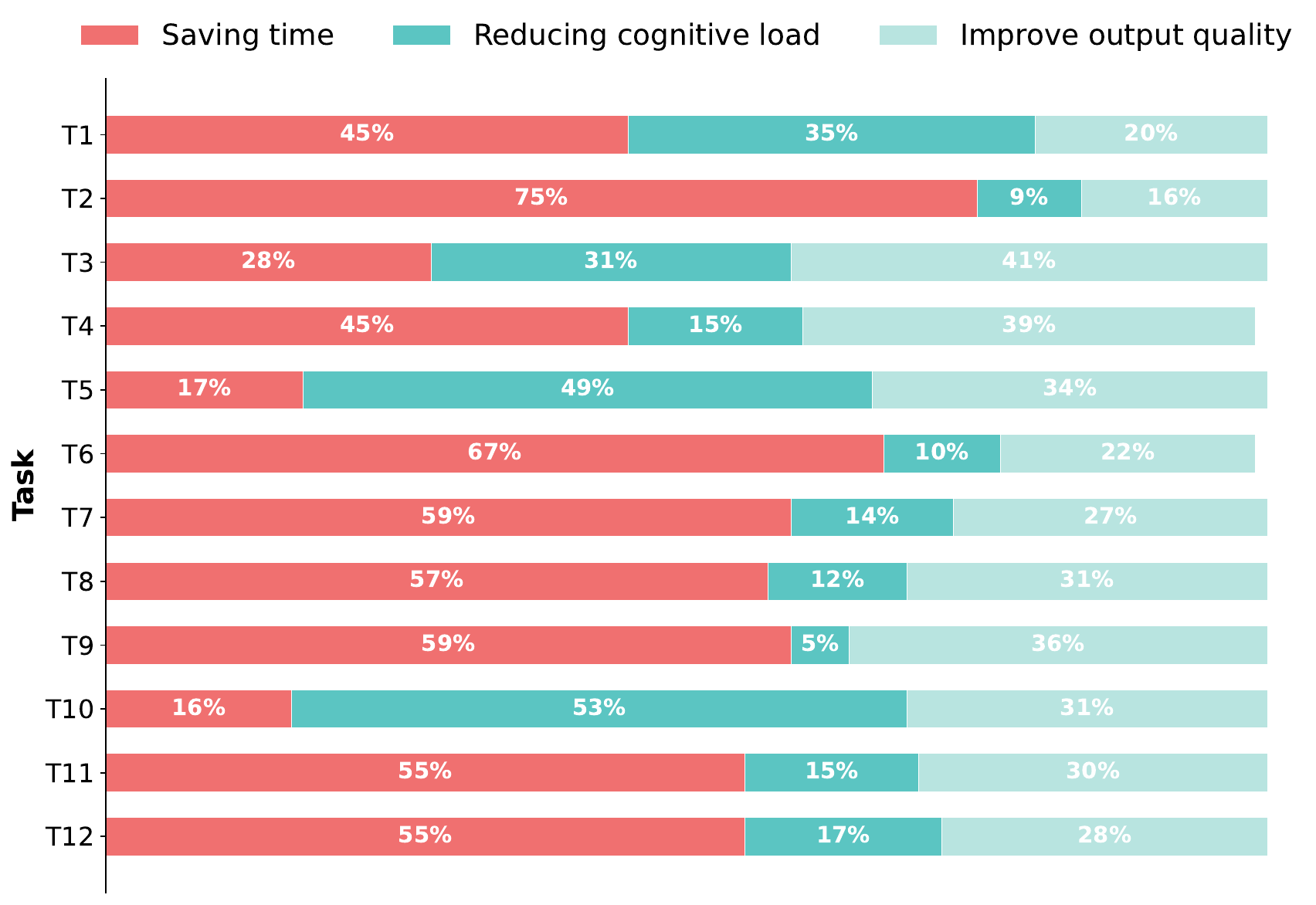}
    \caption{Distribution of students’ reported reasons for using AI across tasks.}
    \label{fig:reason}
\end{minipage}
\hfill
\begin{minipage}[t]{0.48\textwidth}
    \centering
    \includegraphics[width=\linewidth,height=0.29\textheight]{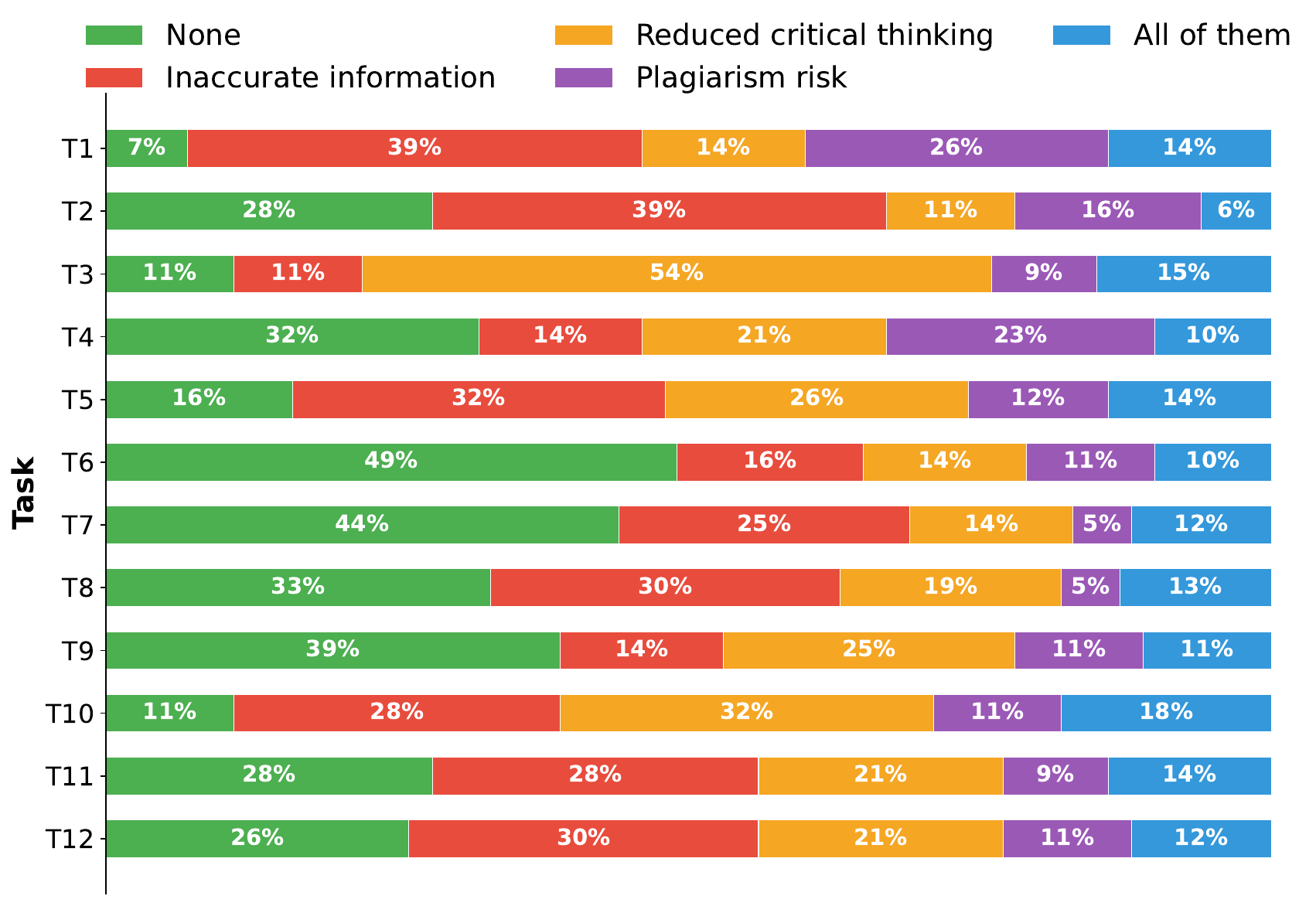}
    \caption{Distribution of students’ reported concerns regarding AI use across tasks. }
    \label{fig:concern}
\end{minipage}
\end{figure*}

Figures~\ref{fig:reason} illustrate the  distributions of students’
reported reasons for using AI, and their associated concerns across twelve academic tasks was illustrated in Figures~\ref{fig:concern}.

\subsubsection{Writing and Revision Tasks}
Writing- and revision-oriented tasks (T1, T4, T9, T11, T12) are strongly dominated by efficiency-related motivations. 
Across this category, \emph{saving time} accounts for the largest share of responses, often exceeding half. 
Students primarily view AI as pragmatic assistance that reduces routine effort rather than reshapes intellectual substance. 
\emph{Improving output quality} is a substantial secondary reason, particularly for revision tasks (T4, T12), where AI enhances clarity and surface-level correctness.

These patterns suggest that students see AI as complementing—not replacing—their writing. 
Because outputs are inspectable and editable, barriers to adoption are low, and students retain control over content and intent.

Concerns are comparatively limited. 
For several tasks (e.g., T9, T11), \emph{None} is the most common response. 
When concerns arise, they focus mainly on \emph{inaccurate information}, outweighing worries about plagiarism or reduced critical thinking. 
Overall, trust in AI writing support appears conditional on students’ ability to verify outputs.

\subsubsection{Ideation and Planning Tasks}
Ideation and planning tasks (T3, T6) show more differentiated patterns. 
For brainstorming (T3), motivations are distributed across \emph{improving output quality} and \emph{reducing cognitive load}, reflecting interest in idea generation and overcoming initial barriers. 
However, \emph{reduced critical thinking} overwhelmingly dominates concerns, indicating awareness of the cognitive risks of delegating creative reasoning. 
Ideation thus represents a sensitive boundary between assistance and substitution.

In contrast, study planning (T6) presents a simpler profile: \emph{saving time} is the dominant reason, and \emph{None} is the most common concern. 
Students appear to view planning as organizational rather than cognitively generative, making AI use less threatening to learning.

\subsubsection{Learning Support Tasks}
Learning support tasks (T7, T8) exhibit highly concentrated reason distributions, with \emph{saving time} clearly dominant and \emph{improving output quality} secondary. 
Students primarily view AI as accelerating access to study materials.

Concerns are minimal: \emph{None} is the largest category, and accuracy- or cognition-related worries are limited. 
Because outputs such as flashcards or resource recommendations can be verified or selectively used, AI is perceived as a low-stakes, supplementary aid.

\subsubsection{Technical and Quantitative Problem-Solving Tasks}
Technical tasks (T5, T10) display the most balanced and complex profiles. 
\emph{Reducing cognitive load} is the leading motivation, with \emph{improving output quality} also prominent, reflecting the complexity and precision required.

At the same time, concerns remain salient. 
\emph{Inaccurate information} and \emph{reduced critical thinking} together account for a substantial share, exceeding other categories—particularly \textbf{inaccurate information}, which is the most frequently reported concern across tasks. 
Students value AI for scaffolding complex reasoning but remain cautious about errors and superficial understanding. 
Compared to writing or general learning tasks, mistakes in technical contexts are harder to detect and potentially more consequential, leading to a more evaluative and risk-sensitive stance toward AI use.

\begin{table*}[t]
\centering
\caption{Students' expected features and design principles for educational AI systems.}
\label{tab:ai-design-principles}

\renewcommand{\arraystretch}{1.3}
\small
\begin{tabular}{
    p{4.2cm}
    p{7.0cm}
    p{5.2cm}
}
\toprule
\textbf{Reason} &
\textbf{Example quotes} &
\textbf{Potential benefits} \\
\midrule

Transparency and verifiability of AI outputs
&
``Always returning a clickable link to a source so it can be easily validated.'' (R10);
``Giving sources for all the information they give.'' (R22);
``AI systems should always provide where it's getting its data from.'' (R42)
&
Students can independently verify AI-generated information and assess its reliability,
reducing blind trust and inappropriate academic use. \\

\midrule

Explicit communication of uncertainty
&
``It should have a percentage next to every response.'' (R23);
``A certainty factor would be an interesting addition.'' (R4);
``More transparency about uncertainty or when guessing is used.'' (R34)
&
Students can better calibrate trust in AI outputs
and decide when additional verification or skepticism is required. \\

\midrule

Explainability and reasoning transparency
&
``More visibility into the inner thought process.'' (R1);
``Clear reasoning steps and transparency about uncertainty.'' (R19);
``More transparent information about the thinking path.'' (R44)
&
Students can critically engage with AI outputs
and treat AI as a cognitive support tool rather than an unquestioned authority. \\

\midrule

Hallucination awareness and error signaling
&
``Flagging hallucinatory content.'' (R6);
``Reduce hallucinations.'' (R8);
``Clear uncertainty indicators and fact-checking.'' (R45)
&
Students can identify potentially unreliable content
and avoid incorporating incorrect information into academic work. \\

\midrule

User feedback and human oversight
&
``Letting you give them feedback, and use that to train in real-time.'' (R38);
``Allow users to flag hallucinated or incorrect information.'' (R53)
&
Supports human-in-the-loop interaction
and increases accountability in educational AI systems. \\

\midrule

Skepticism toward AI use
&
``Features that prioritize perceived trustworthiness undermine nuance of AI deception.'' (R11)
&
Highlights the need for AI systems that respect skepticism
rather than attempting to enforce trust or authority. \\

\bottomrule
\end{tabular}
\end{table*}

\subsection{RQ3: Expected System Features and Design Principles of AI System}
After participants reported their primary reasons and concerns regarding AI use across academic tasks, we administered a follow-up survey to collect open-ended responses about the design features they expected AI systems to provide to address the most prevalent concerns across tasks: inaccurate information.
This resulted in 53 valid responses ($N = 53$) out of the 57 participants, 
in which respondents described the features and design principles they expect AI systems to provide 
to address concerns about inaccurate information and hallucinations. 
We refer to individual respondents as R1--R53 throughout this section.

As Table 3 shows, a dominant theme across participants' open-ended responses concerns the need for
\textbf{transparency and verifiability in AI-generated outputs}.
Many participants emphasized that AI systems should consistently provide
explicit source citations, clickable references, or other mechanisms
that allow users to trace where information originates.
Rather than treating AI outputs as authoritative answers,
participants repeatedly framed trustworthy AI as a system that enables
independent verification.
As one participant explained,
``Always returning a clickable link to a source so it can be easily validated'' (R10).
Similarly, another participant stated,
``I think AI systems should always provide where it's getting its data from
so the user is able to check it on their own'' (R42).

Closely related to source transparency is the expectation that AI systems should
\textbf{explicitly communicate uncertainty}.
Participants frequently requested confidence scores, reliability metrics,
or clear indicators when the system is unsure or unable to answer.
Several participants suggested that AI should not attempt to mask uncertainty
through fluent or confident language.
For example, one participant noted,
``It should have a percentage next to every response.
That percentage should tell how factual AI thinks the information that it presented is'' (R23),
while another stated,
``I think I would like if it just shares when it's not able to answer'' (R2).
Importantly, some participants cautioned that confident presentation
can be misleading in educational settings.
As R11 argued,
``Features that prioritize perceived trustworthiness undermine the nuance of AI deception,''
emphasizing that productive AI use requires sustained skepticism rather than blind trust.

\textbf{Explainability and reasoning transparency} also emerged as a central design consideration.
Participants expressed interest in greater visibility into how AI systems generate responses,
including access to reasoning steps, intermediate logic,
or the ability to intervene during response generation.
One participant requested
``more visibility into the inner thought process, more ability to do surgery
on previous responses in a conversation'' (R1),
while another highlighted the importance of
``clear reasoning steps, transparency about uncertainty,
and real-time fact-checking'' (R19).
Several participants framed explainability not as a way to increase trust,
but as a means of supporting critical engagement.
For example, a participant described valuing systems that
``show the steps of their thinking where I can intervene in the middle
and reshape the direction.''

Participants also raised strong concerns about
\textbf{hallucinations and misleading outputs},
underscoring the need for explicit error signaling and safeguards.
Many respondents requested features that flag potentially hallucinated content,
provide warnings, or incorporate built-in fact-checking mechanisms.
As one participant succinctly stated,
``Flagging hallucinatory content.
Proper warnings (not in fine print) also help'' (R6).
Others emphasized reducing hallucinations altogether,
with one participant stating,
``Reduce hallucinations'' (R8),
and another calling for
``clear uncertainty indicators, citations linked to verifiable sources,
and explanations of reasoning steps'' to increase trust (R45).

\section{Discussion}

\subsection{Conditional Automation and the Limits of Adoption}

Findings from RQ1 show a consistent gap between students’ desired levels of automation
and their reported AI usage across most academic tasks.
Importantly, this gap does not reflect general resistance to AI,
but rather a pattern of \emph{conditional adoption}.
Students express openness to greater AI involvement
while simultaneously withholding full reliance in practice.

This conditionality is strongly shaped by task characteristics.
Tasks that are procedural, supportive, or easily verifiable
(e.g., writing revision, summarization, citation formatting, and debugging code)
tend to cluster in regions of both high desired automation and high usage.
In contrast, tasks that are socially sensitive
(e.g., drafting emails) or cognitively generative
(e.g., brainstorming, quantitative reasoning)
exhibit larger gaps, with desired automation exceeding actual usage.
These patterns suggest that students do not evaluate AI
solely based on its technical capability,
but also based on contextual factors such as accountability,
learning value, and error detectability.

\subsection{Task-Specific Motivations and Risk Awareness}

Results from RQ2 provide insight into why such conditional adoption emerges.
Across tasks, efficiency-related motivations—particularly saving time
and reducing cognitive load—dominate students’ reported reasons for using AI.
However, these motivations are not applied uniformly.
Instead, they interact with task-specific concerns
that reflect students’ awareness of cognitive and epistemic risks.

For writing, revision, and learning support tasks,
AI is primarily conceptualized as a pragmatic assistant.
These tasks are characterized by outputs that are inspectable,
editable, and low risk,
which helps explain why concerns are minimal
and why “None” frequently emerges as the dominant concern.
In these contexts, students appear comfortable leveraging AI
to streamline effort without feeling that core learning is compromised.

In contrast, ideation and technical problem-solving tasks
activate distinct concern profiles.
For brainstorming, concerns about reduced critical thinking
dominate overwhelmingly,
suggesting that students view ideation as central
to intellectual ownership and learning.
For technical and quantitative tasks,
concerns about inaccurate information and shallow understanding
feature prominently,
reflecting the higher stakes associated with errors
and the difficulty of verification.
These findings demonstrate that students’ concerns
are not abstract fears about AI,
but grounded assessments of how AI assistance
intersects with task goals and learning outcomes.

\subsection{Design Expectations as Responses to Identified Risks}

Findings from RQ3 reveal that students’ expectations for AI system features
directly respond to the tensions identified in RQ1 and RQ2.
Rather than requesting more automation,
participants consistently called for design features
that support \emph{verification, reflection, and skepticism}.

Transparency and verifiability emerged as the most dominant design expectation.
Students repeatedly emphasized the need for source citations,
clickable references, and traceable evidence,
particularly in response to concerns about inaccurate information.
These requests align closely with the elevated concern profiles
observed in technical and writing-related tasks,
where correctness is essential and errors may propagate easily.

Similarly, explicit communication of uncertainty—through confidence scores,
reliability indicators, or admission of inability to answer—
addresses students’ reluctance to over-rely on AI
in cognitively demanding tasks.
Rather than equating confidence with trust,
students explicitly rejected persuasive or authoritative presentation,
arguing instead for systems that make uncertainty visible.
This stance reflects a mature understanding of AI limitations
and positions skepticism as a productive component
of educational AI use.

\section{Conclusion and Future Work}

In this paper, we examined how graduate CS students engage with AI systems
across a range of academic tasks,
focusing on desired automation levels,
actual usage,
underlying motivations,
perceived concerns,
and expectations for AI system design.
Through a mixed-methods user study,
we showed that students’ adoption of AI is highly task dependent,
with consistent gaps between desired and actual automation
that reflect cautious, evaluative decision-making rather than indiscriminate use.
We further demonstrated that efficiency-driven motivations
coexist with task-specific concerns,
particularly around inaccurate information and reduced critical thinking.
Finally, we identified key design expectations for educational AI systems,
including transparency, verifiability, uncertainty communication,
and support for user skepticism.
Together, our findings provide empirical grounding and design-relevant insights
for building educational AI systems
that align with students’ learning goals
and preserve human agency in student–AI collaboration. In future work, this study can be extended to include a broader range of academic tasks and a more diverse student population beyond computer science students. Furthermore, it would be beneficial to design the survey with open-ended questions for capturing students’ concerns and reasons for using AI. Open-ended responses would allow participants to articulate more nuanced, context-dependent perspectives that may not fit within predefined categories.


\bibliographystyle{ACM-Reference-Format}
\bibliography{main}

\appendix

\section{Full Participant List}
\label{app:participants_full}

\onecolumn
\small

\begin{longtable}{p{0.8cm}p{2.2cm}p{5.2cm}p{4.2cm}}
\caption{Participant-level demographics of OMSCS students who reported AI tool use in coursework (N=57).}\\
\toprule
\textbf{ID} & \textbf{Age Group} & \textbf{OMSCS Specialization} & \textbf{AI Usage Frequency} \\
\midrule
\endfirsthead

\toprule
\textbf{ID} & \textbf{Age Group} & \textbf{OMSCS Specialization} & \textbf{AI Usage Frequency} \\
\midrule
\endhead

P1 & 30--40 & Human--Computer Interaction & Several times per week \\
P2 & 25--30 & Machine Learning & Daily or almost daily \\
P3 & 25--30 & Human--Computer Interaction & Often \\
P4 & 18--24 & Artificial Intelligence & Often \\
P5 & 30--40 & Human--Computer Interaction & Several times per week \\
P6 & 25--30 & Human--Computer Interaction & Rarely \\
P7 & 30--40 & Artificial Intelligence & Several times per week \\
P8 & Above 40 & Computing Systems & Several times per week \\
P9 & 30--40 & Artificial Intelligence & Daily or almost daily \\
P10 & 18--24 & Artificial Intelligence & Several times per week \\
P11 & 18--24 & Not specified & Daily or almost daily \\
P12 & 18--24 & Artificial Intelligence & Daily or almost daily \\
P13 & 18--24 & Undecided & Often \\
P14 & 30--40 & Undecided & Daily or almost daily \\
P15 & 30--40 & Human--Computer Interaction & Daily or almost daily \\
P16 & 30--40 & Computing Systems & Rarely \\
P17 & 18--24 & Undecided & Several times per week \\
P18 & 30--40 & Computing Systems & Often \\
P19 & 25--30 & Artificial Intelligence & Often \\
P20 & 30--40 & Human--Computer Interaction & Several times per week \\
P21 & 25--30 & Computing Systems & Several times per week \\
P22 & 18--24 & Human--Computer Interaction & Rarely \\
P23 & 18--24 & Human--Computer Interaction & Several times per week \\
P24 & 30--40 & Artificial Intelligence & Daily or almost daily \\
P25 & 18--24 & Artificial Intelligence & Often \\
P26 & 25--30 & Human--Computer Interaction & Rarely \\
P27 & 18--24 & Human--Computer Interaction & Rarely \\
P28 & 18--24 & Human--Computer Interaction & Rarely \\
P29 & 30--40 & Machine Learning & Daily or almost daily \\
P30 & 25--30 & Human--Computer Interaction & Rarely \\
P31 & 25--30 & Human--Computer Interaction & Daily or almost daily \\
P32 & 25--30 & Computing Systems & Daily or almost daily \\
P33 & 30--40 & Artificial Intelligence & Daily or almost daily \\
P34 & 18--24 & Machine Learning & Daily or almost daily \\
P35 & 25--30 & Human--Computer Interaction & Never \\
P36 & 30--40 & Artificial Intelligence & Never \\
P37 & 18--24 & Artificial Intelligence / Machine Learning & Several times per week \\
P38 & Above 40 & Artificial Intelligence & Daily or almost daily \\
P39 & 18--24 & Undecided & Several times per week \\
P40 & 18--24 & Human--Computer Interaction & Rarely \\
P41 & 25--30 & Artificial Intelligence & Daily or almost daily \\
P42 & 18--24 & Artificial Intelligence & Daily or almost daily \\
P43 & 18--24 & Not specified & Daily or almost daily \\
P44 & 25--30 & Human--Computer Interaction & Rarely \\
P45 & 25--30 & Human--Computer Interaction & Several times per week \\
P46 & 18--24 & Human--Computer Interaction & Rarely \\
P47 & 18--24 & Artificial Intelligence & Often \\
P48 & 18--24 & Artificial Intelligence & Often \\
P49 & 30--40 & Computing System & Daily or almost daily \\
P50 & 25--30 & Human--Computer Interaction & Daily or almost daily \\
P51 & 25--30 & Human--Computer Interaction & Several times per week \\
P52 & Above 40 & Machine Learning & Daily or almost daily \\
P53 & Above 40 & Human--Computer Interaction & Rarely \\
P54 & 18--24 & Artificial Intelligence & Several times per week \\
P55 & 25--30 & Human--Computer Interaction & Several times per week \\
P56 & 25--30 & Human--Computer Interaction & Daily or almost daily \\
P57 & 30--40 & Artificial Intelligence & Daily or almost daily \\

\bottomrule
\end{longtable}

\twocolumn

\end{document}